 \documentclass[prx,showpacs,superscriptaddress]{revtex4}
\usepackage{graphicx}
\usepackage{mathrsfs}
\usepackage{dcolumn}
\usepackage{bm}
\usepackage{amsmath}
\usepackage{amsfonts}
\usepackage{color}

\usepackage{setspace}

\usepackage{amssymb}
\usepackage{amsthm}
\usepackage{amsfonts}
\usepackage{listings}
\usepackage{enumerate}
\usepackage{latexsym}
\usepackage{color}
\usepackage{bm}
\usepackage{hyperref}
\hypersetup{
 pdfnewwindow=true, colorlinks=true,
 linkcolor=blue, anchorcolor=blue,
 citecolor=blue, filecolor=blue,
 menucolor=blue, urlcolor=blue}

\usepackage{psfrag}

\usepackage{bm}
\usepackage{graphicx}
\usepackage{subfigure}

\RequirePackage[normalem]{ulem} 
\RequirePackage{color}\definecolor{RED}{rgb}{1,0,0}\definecolor{BLUE}{rgb}{0,0,1} 

\input{epsf}

\begin{document}

\tolerance 10000

\newcommand{\vk}{{\bf k}}

\draft

\title{Topological Electronic States in HfRuP Family Superconductors}
\author{Yuting Qian}
\thanks{These authors contributed equally to this work.}
\affiliation{Beijing National Laboratory for Condensed Matter Physics,
and Institute of Physics, Chinese Academy of Sciences, Beijing 100190, China}
\affiliation{University of Chinese Academy of Sciences, Beijing 100049, China}

\author{Simin Nie}
\thanks{These authors contributed equally to this work.}
\affiliation{Beijing National Laboratory for Condensed Matter Physics,
and Institute of Physics, Chinese Academy of Sciences, Beijing 100190, China}
\affiliation{Department of Materials Science and Engineering, Stanford University, Stanford, California 94305, USA}

\author{Changjiang Yi}
\thanks{These authors contributed equally to this work.}
\affiliation{Beijing National Laboratory for Condensed Matter Physics,
and Institute of Physics, Chinese Academy of Sciences, Beijing 100190, China}

\author{Lingyuan Kong}
\affiliation{Beijing National Laboratory for Condensed Matter Physics,
and Institute of Physics, Chinese Academy of Sciences, Beijing 100190, China}
\affiliation{University of Chinese Academy of Sciences, Beijing 100049, China}

\author{Chen Fang}
\affiliation{Beijing National Laboratory for Condensed Matter Physics,
and Institute of Physics, Chinese Academy of Sciences, Beijing 100190, China}
\affiliation{University of Chinese Academy of Sciences, Beijing 100049, China}
\affiliation{CAS Center for Excellence in Topological Quantum Computation, University of Chinese Academy of Sciences, Beijing 100190, China}
\affiliation{Songshan Lake Materials Laboratory, Dongguan, Guangdong 523808, China} 

\author{Tian Qian}
\affiliation{Beijing National Laboratory for Condensed Matter Physics,
and Institute of Physics, Chinese Academy of Sciences, Beijing 100190, China}
\affiliation{Songshan Lake Materials Laboratory, Dongguan, Guangdong 523808, China} 

\author{Hong Ding}
\affiliation{Beijing National Laboratory for Condensed Matter Physics,
and Institute of Physics, Chinese Academy of Sciences, Beijing 100190, China}
\affiliation{CAS Center for Excellence in Topological Quantum Computation, University of Chinese Academy of Sciences, Beijing 100190, China}
\affiliation{Songshan Lake Materials Laboratory, Dongguan, Guangdong 523808, China} 

\author{Youguo Shi}
\affiliation{Beijing National Laboratory for Condensed Matter Physics,
and Institute of Physics, Chinese Academy of Sciences, Beijing 100190, China}
\affiliation{University of Chinese Academy of Sciences, Beijing 100049, China}
\affiliation{Songshan Lake Materials Laboratory, Dongguan, Guangdong 523808, China} 

\author{Zhijun Wang}
\email{wzj@iphy.ac.cn}
\affiliation{Beijing National Laboratory for Condensed Matter Physics,
and Institute of Physics, Chinese Academy of Sciences, Beijing 100190, China}
\affiliation{University of Chinese Academy of Sciences, Beijing 100049, China}

\author{Hongming Weng}
\email{hmweng@iphy.ac.cn}
\affiliation{Beijing National Laboratory for Condensed Matter Physics,
and Institute of Physics, Chinese Academy of Sciences, Beijing 100190, China}
\affiliation{University of Chinese Academy of Sciences, Beijing 100049, China}
\affiliation{CAS Center for Excellence in Topological Quantum Computation, University of Chinese Academy of Sciences, Beijing 100190, China}
\affiliation{Songshan Lake Materials Laboratory, Dongguan, Guangdong 523808, China} 

\author{Zhong Fang}
\affiliation{Beijing National Laboratory for Condensed Matter Physics,
and Institute of Physics, Chinese Academy of Sciences, Beijing 100190, China}
\affiliation{University of Chinese Academy of Sciences, Beijing 100049, China}
\date{\today}

\begin{abstract}
Based on the first-principles calculations and experimental measurements, we report that the hexagonal phase of ternary transition metal pnictides TT'X (T=Zr, Hf; T'=Ru; X=P, As), which are well-known noncentrosymmetric superconductors with relatively high transition temperatures, host nontrivial bulk topology. Before the superconducting phase transition, we find that HfRuP belongs to a Weyl semimetal phase with 12 pairs of type-II Weyl points, while ZrRuAs, ZrRuP and HfRuAs belong to a topological crystalline insulating phase with trivial Fu-Kane $\mathbb Z_2$ indices but \emph{nontrivial} mirror Chern numbers. High-quality single crystal samples of the noncentrosymmetric superconductors with these two different topological states have been obtained and the superconductivity is verified experimentally. The wide-range band structures of ZrRuAs have been identified by ARPES and reproduced by theoretical calculations. Combined with intrinsic superconductivity, the nontrivial topology of the normal state may generate unconventional superconductivity in both bulk and surfaces. Our findings could largely inspire the experimental searching for possible topological superconductivity in these compounds.
\end{abstract}

\maketitle
\section*{INTRODUCTION}

Topological insulators (TIs)~\cite{kane2010,qi2011} and semimetals~\cite{wan2011topological,xu2011chern,wang2012dirac,wang2013three,weng2015weyl} have received a tremendous amount of attention in the last decade due to the appearance of exotic properties, such as spin-momentum locked gapless surface state in TIs~\cite{PhysRevLett.106.257004,zhang2013spin}, Fermi-arc states~\cite{xu2016observation,xu2015discovery,wang2016observation2} and negative magnetoresistance in Weyl semimetals (WSMs)~\cite{weng2015weyl,huang2015observation,zhang2016signatures}. These insulators can be characterized by topological invariants/indices, like Fu-Kane $\mathbb Z_2$ indices~\cite{PhysRevLett.98.106803} and mirror Chern numbers~\cite{hsieh2012topological,nie2016band} for TIs and topological crystalline insulators (TCIs), respectively. However, WSMs are topological metallic states with discrete accidental twofold degenerate points, described by the three-dimensional (3D) Weyl equation. Due to the lack of strict Lorentz invariance, the type-II Weyl points can be strongly tilted~\cite{soluyanov2015type}, which have no analogy in high-energy physics. In contrast to the point-like bulk Fermi surfaces of the type-I WSMs~\cite{weng2015weyl,huang2015weyl,lv2015experimental,lv2015observation,lv2015observation2,nie2017topological,nie2019magnetic,PhysRevLett.117.236401}, these type-II WSMs~\cite{deng2016experimental,jiang2017signature,tamai2016fermi,liang2016electronic,wang2016mote} have both electron pockets and hole pockets touching at the Weyl points, resulting in various novel physical properties~\cite{kumar2017extremely,shekhar2015extremely}.

Topological materials that host superconductivity are ideal systems to detect topological superconductivity (TSC) and Majorana fermions~\cite{yan2013large,wu2015,Schoop2015,chang2016,wang2016spontaneous,xie2017,nie2018}. The topological surface Dirac-cone states can be used to generate two-dimensional (2D) TSC induced by the intrinsic bulk superconductivity ~\cite{Fu2008superconducting,Fu2010odd,alicea2012new,sato2017topological}. Very recently, the superconducting gap of the predicted topological surface Dirac-cone states in FeTe$_{1-x}$Se$_x$~\cite{wang2015,xu2016} has been detected in recent angle-resolved photoemission spectroscopy (ARPES)~\cite{Zhang182} and scanning tunneling microscope experiments~\cite{Wang333}. In noncentrosymmetric WSMs, 3D time-reversal symmetric TSC can arise from sign-changing superconductivity in Fermi surfaces with different Chern numbers \cite{qi2010,Hosur204}. However, to the best of our knowledge, almost all noncentrosymmetric WSMs need external pressure or doping to induce or enhance superconductivity~ \cite{pan2015pressure,kang2015superconductivity,qi2016superconductivity,chen2016superconductivity,li2017concurrence,xu2019topological}. Due to the lack of suitable candidates, 3D TSC is studied very little experimentally. Therefore, the material proposal of a WSM with high-quality single crystals and relatively higher superconducting transition temperature (T$_C$) is of great interest.

Ternary transition metal pnictides TT'X (T=Zr, Hf; T'=Ru; X=P, As) are a series of well-known superconductors~\cite{barz1980ternary,meisner1983superconductivity}. As we know, there are three different types of crystal structures for these compounds \cite{MULLER1983177,meisner1983superconductivity}, {\it i.e.} the Fe$_2$P-type hexagonal structure (h-phase), the TiNiSi-type orthorhombic structure (o-phase), and the TiFeSi-type orthorhombic structure (o$'$-phase). Superconductivity is found in both h- and o-phases, and the superconducting transition temperature $T_C$ is generally higher for h-phase than that for o-phase. In this work, we only focus on the h-phase of TT'X, exhibiting relatively high $T_C$ (such as 12.7 K for HfRuP \cite{barz1980ternary}, 13.3 K for ZrRuP and 12 K for ZrRuAs \cite{meisner1983superconductivity}). Based on the first-principles calculations, the nontrivial topological properties of these materials in the normal state (above T$_C$) are revealed. 
When spin-orbit coupling (SOC) is ignored, they possess two nodal rings slightly above the Fermi energy (E$_F$) in the $k_z=0$ plane, with each surrounding a K point in the hexagonal Brillouin zone (BZ). After the consideration of SOC, they enter \emph{either} a WSM phase ({\it e.g.} HfRuP) with 12 pairs of type-II Weyl points (WPs) due to the lack of inversion symmetry, \emph{or} a topological crystalline insulating (TCI) phase ({\it e.g.} ZrRuAs) with trivial Fu-Kane $\mathbb Z_2$ indices \cite{Fu2007topo} but \emph{nontrival} mirror Chern numbers. The nontrivial electronic topology in these materials could intrigue tremendous experimental study of the interplay between topological electronic states and superconductivity.

\section*{RESULTS AND DISCUSSION}
\subsection{Crystal structure and electronic band structures}
The h-phase of TT'X is of space group $P\bar 6 2m$ (\#189) with a layered structure. Each layer in the hexagonal lattice is occupied by
either T and X atoms or T' and X atoms. All atoms have positions in the layers parallel to the crystallographic $ab$-plane and separated by a half of the lattice constant $c$. The triangular clusters of three T' atoms (T'$_3$) are formed in the $ab$-plane. In the crystal structure of the TT'X shown in Figs.~1(a) and (b), the T'$_3$ clusters and the planer structure are clearly shown. The high-symmetry $k$-points and surface projections are shown in Fig.~1(c). The structure has two kinds of mirror symmetries, $m_z$ and $m_x$, which are vital to define the mirror Chern numbers as will be shown below.  Meantime, we have successfully grown the single crystals of ZrRuAs and HfRuP, as shown in Figs. 1(d) and (e), respectively. The hexagonal structure and superconductivity are confirmed by the x-ray diffraction (XRD) and resistivity measurements, respectively. More details and data ({\it i.e.} magnetic susceptibility) can be found in the Supplementary Material.

We first checked the electronic band structure without SOC. Among these compounds, we mainly investigated HfRuP and ZrRuAs for details in the following, as typical examples of the type-II WSM phase and TCI phase, respectively. More results on other compounds are presented in the Supplementary Material. From the band dispersion of HfRuP in Fig. 2(a), one can notice there is a direct energy gap shadowed in light blue near E$_F$, except the band crossings along both $M-K$ and $K-\Gamma$ lines. These two lines are actually in the $k_z = 0$ plane, where the $m_z$ symmetry is present. The $m_z$ eigenvalues of the two crossing bands are computed to be $\pm 1$, respectively. Thus, the two crossing points are parts of the $m_z$-protected nodal rings, each of which surrounds a K point in the $k_z = 0$ plane, as depicted in Fig.~2(c).
This situation is different from that in CaAgAs \cite{yamakage2015line}, where there is only one nodal ring surrounding the $\Gamma$ point.
The two nodal rings circled around two K points are also found for all other compounds (see electronic band structures of ZrRuAs, HfRuAs and ZrRuP in Supplementary Section A).
We conclude that the band inversion happens at the K point, which is supported by the theory of topological quantum chemistry~\cite{tqc2017,wang2019}. By exchanging the highest valence band ($\Gamma_4$) and the lowest conduction band ($\Gamma_1$) at the K point (with little group $D_{3h}$) only, the occupied bands become trivial, being a linear combination of elemental band representations~\cite{tqc2017}.

After including SOC, the band structure doesn't change too much, but the bands do split due to the lack of inversion symmetry.
To confirm the reliability of the density functional theory (DFT) band structures, we have performed ARPES measurement for ZrRuAs, shown in Figs.~2(f)-(i). The observed spectra along H-K-H and L-M-L lines match very well with the DFT calculations (red lines in Figs.~2(g) and (i)), especially for the low-energy bands. We clearly see that the energy bands at K point are much lower than that at M point. In addition, the degeneracy of the two nodal rings is lifted by SOC. The 2D time-reversal invariant (TRI) planes ({\it e.g.} $k_y = 0$ and $k_z = 0$ etc.) become fully gapped, making the $\mathbb Z_2$ invariants well-defined. In CaAgAs, the single nodal line enclosing $\Gamma$ point guarantees that the $k_y = 0$ (or $k_z = 0$) plane is $\mathbb Z_2$ nontrivial with an infinitesimal SOC gap. But, it's not the case with two nodal rings around two K points. The $\mathbb Z_2$ invariants for both $k_y = 0$ and $k_z = 0$ planes remain trivial in the series of these compounds. Note that the $k_y = 0$ plane is gapped even without SOC, and no gapless point is found in all TRI planes. To confirm triviality of $\mathbb Z_2$ invariants, we have calculated the Wannier charge centers (WCCs) of the $k_z$-directed ($k_y$-directed) Wilson loops as a function of $k_x$ (called Wilson-loop bands) for the $k_y = 0$ ($k_z = 0$) plane. The results of ZrRuAs are shown in Figs.~3(a) and (b), suggesting a trivial $\mathbb Z_2$ invariant in the $k_y = 0$ plane and $k_z = 0$ plane. Accordingly, the Fu-Kane $\mathbb Z_2$ indices~\cite{Fu2007topo} for the 3D bulk are computed to be (0;000) for all the compounds. The detailed calculations for the $\mathbb Z_2$ invariants in all six TRI planes (only four of them are distinct due to the symmetry) are presented in Supplementary Section B.
Furthermore, the symmetry indicators~\cite{ashvin2017,song2017,Jorrit2017,zhang2019,wanxg2019} for these compounds are computed to be $\mathbb Z_{3m,0}=1$ and $\mathbb Z_{3m,\pi}=0$, revealing the topological nature of the SOC gap (shadowed in light blue) in Fig.~2(b).

\subsection{Mirror Chern numbers and WPs}

Due to the presence of mirror symmetry, the mirror Chern number is well defined as long as the mirror plane is fully gapped. Because time-reversal symmetry commutes with the mirror symmetries, the Chern numbers satisfy $C_i =  -C_{-i}$, with the subscript $\pm i$ representing the mirror eigenvalues in the presence of SOC. With time reversal symmetry, the mirror Chern number is defined as
$C_m = (C_{+i}-C_{-i})/2$. As we know, it can be further reduced to $\chi_{+i}-\chi_{-i}$ in \emph{half} of the mirror plane, where  $\chi_{+i(-i)}$ is easily obtained in the plot of Wilson-loop bands, by counting the number of the positively-sloped bands crossing a horizontal reference line [the dashed line in Figs.~3(c) or (d)] and subtracting from it the number of the negatively-sloped crossing ones in the mirror eigenvalue $+i~(-i)$ subspace. The results of ZrRuAs for the $k_z = 0$ plane are shown in Fig.~3(b). The mirror Chern number $C_{m_z}$ in ZrRuAs is computed to be $-2$ for the $k_z$ = 0 plane, while it's zero for the $k_z =\pi$ plane. That's the case for all the compounds [see more in Supplementary Section C].

The lack of inversion symmetry allows the appearance of WPs in the systems. The nonzero mirror Chern number $C_{m_z} =-2$ suggests there are some strings of gauge singularities ({\it i.e.} the Dirac string) going through the $k_z = 0$ plane \cite{bernevig2015s}, which have to either terminate at WPs in the 3D BZ, or thread some other nontrivial planes. Our systematic calculations show that these compounds can be classified into two classes: i) one has zero Chern number $C_{m_x} =0$ with 12 pairs of type-II WPs, termed a WSM phase; ii) the other one has \emph{nonzero} mirror Chern number $C_{m_x} =2$ with no WPs, termed a TCI phase. In the WCCs of HfRuP for the $k_x=0$ plane in Fig.~3(c), $C_{m_x}$ is obtained to be 0. WPs are found in this compound, which is consistent with the topological WSM phase. However, for ZrRuAs, HfRuAs and ZrRuP, $C_{m_x}$ turns out to be $2$, as shown in Fig. 3(d) and in Supplementary Section C. Accordingly, no WP is found in these three materials. The detailed calculations of mirror Chern numbers are presented in Supplementary Section C.

By checking the energy gap and the associated topological monopole charge, we find that 6 pairs of WPs emerge from each nodal ring. Thus, there are 12 pairs of WPs in total (as shown in the first BZ in Fig.~2(c)). They reside at the same energy, because they are all related by either time-reversal symmetry or the crystalline symmetry $D_{3h}$ (including 12 symmetry operators). The coordinate of the WP W1 enclosed by a dashed circle in Fig.~2(c) is [0.2761$\bf a$*, $-$0.4654$\bf b$*, 0.02439$\bf c$*]. From the band dispersion of the WP W1 along the $k_z$-direction [Fig.~2(d)] and the P-Q direction [Fig.~2(e)], we conclude that it belongs to a type-II WP~\cite{xu2015}, and its energy level (E$_W$) is about 28 meV above E$_F$ ({\it i.e.} E$_W$-E$_F= 28$ meV), very close to the Fermi energy. The topological monopole charge is computed with the Wilson-loop method applied on an enclosed manifold surrounding a single WP. The monopole charge of the WP W1 is +1. The distribution of all the WPs above the
$k_z = 0$ plane is illustrated in Fig.~2(c), with the ``+(o)" symbol representing the topological monopole charge of $+1(-1)$, while these below the $k_z = 0$ plane possess the opposite monopole charge shown in Fig.~2(c) because of the $m_z$ symmetry.

\subsection{Fermi arcs on surfaces}

Surface Fermi arcs connecting the projections of two WPs with opposite chirality are expected in a WSM. For this purpose, the surface spectrum is computed based on the surface Green's function method \cite{Sancho_1985} in the maximally localized Wannier function (MLWF) Hamiltonian of a half-infinite structure. First, the (001)-surface energy contour of HfRuP is obtained in Fig.~4(a) with $E-E_W=0$ meV. Since the two WPs with opposite chirality project onto the same point on the (001)-surface, no topological arc states are guaranteed to come out from the projections. However, we find that there are two trivial arc states (following the two dashed guiding lines) coming out of each WP projection: one is crossing the $k_x = 0$ line; the other one is crossing the BZ boundary, ({\it i.e.} the $\bar K- \bar M$ line. Second, the computed (100)-surface energy contour is presented in Fig.~4(b). Since they are type-II WPs, the WPs should be located at the touching points between the electron pocket and the hole pocket. We do see that the projected points A and B are the touching points of two pockets. The constant energy contour with energy slightly below (or above) E$_{W}$ is presented in Supplementary Section D. We find that two surface Fermi arc states (indicated by dashed lines) are connected to the projected points ({\it i.e.} A and B). For the projected point C, it's hard to see any surface state from it, because it is not projected onto a proper surface. At last, the computed (010)-surface energy contour is obtained and shown in Fig.~4(c), where the WPs with the same chirality project onto each other. As long as the projected electron/hole pockets (enclosing the projections of the WPs) are separated from each other, two topological Fermi arc states can be expected. Unfortunately, the metallic bulk states are projected into a big continuum, which makes the Fermi arc states invisible. Our ARPES experiment to search for the arc states on the (100)-surface is still in process.

\subsection{Exotic TSC}
With intrinsic superconductivity, topological materials are promising platforms to realize TSCs owing to the nontrivial topology of the wave function in normal states. For example, surface Dirac fermions can realize a 2D TSC even for an $s$-wave pairing state~\cite{Fu2008superconducting,wang2015}. Also, the FS topology in the normal state directly affects the TSC. In 3D, the integer topological quantum number in a time-reversal invariant superconductor is determined by the sign of the pairing order parameter and the first Chern number of the Berry phase gauge field on the FSs~\cite{qi2011}. A WSM phase of a superconductor hosts the nontrivial FSs originating from WPs. Providing the two key ingredients: the nontrivial FSs and superconductivity, the WSM phase of HfRuP can be served as a good platform to realize the TRI TSC in 3D. Besides, the previous works~\cite{shingo2015,ueno2013} report that the nontrivial mirror Chern number can generate multiple Majorana fermions in Cd$_3$As$_2$ and SrRuO$_4$. These compounds in the TCI phase are very promising candidates to search for the topological crystalline superconductors too.

\section*{Discussion}
Based on the DFT calculations, we find that there are two nodal rings in the band structure without
SOC for the h-phase of TT'X, which is different from the situation in CaAgAs. After including SOC, they enter
either a WSM phase with 12 pairs of type-II WPs, or a TCI phase with non-zero mirror Chern numbers. The single
crystals for the two distinct topological phases are grown successfully. Their superconductivity and
electronic band structures are verified by our resistivity, magnetic susceptibility and ARPES measurements, respectively.
The series of Ru-based compounds are the desired single-crystal materials, which host both superconductivity below T$_C$ and
topological states above T$_C$. The experimental study of the interplay between superconductivity and Weyl or nontrivial mirror Chern states would be stimulated after this work.

At the stage of finalizing the present paper, we are aware of the mention of Weyl nodes in similar materials in the Ref.\cite{ivanov2019}. Our result of WPs in HfRuP is consistent with it, while the result for ZrRuP is different. That is because the topology (the annihilation of Weyl points) in ZrRuP may be sensitive to the parameters of the structure.

\section*{METHODS}
The first-principles calculations were performed based on the DFT with the projector augmented wave (PAW) method \cite{paw1,paw2} as implemented in VASP package \cite{KRESSE199615,vasp}. The generalized gradient approximation (GGA) of Perdew-Burke-Ernzerhof (PBE) type was adopted for the exchange-correlation functional \cite{pbe}. The kinetic energy cutoff of the plane wave basis was set to 400 eV. A 10 $\times$10 $\times$ 16 k-point mesh for BZ sampling was adopted. The experimental lattice parameters were employed \cite{Meisner1983,MEISNER1983983}. The internal atomic positions were fully relaxed until the forces on all atoms were smaller than 0.01 eV/\AA~[the relaxed atomic positions are shown in Supplementary Section A]. The electronic structures were carried out both with and without SOC. The topological invariants and chiral charge were computed through the Wilson-loop technique. The MLWF method was used to calculate the surface states \cite{mlwf}.


\section*{DATA AVAILABILITY}
The data that support the findings of this study are available from the corresponding author upon reasonable request.

\section*{ACKNOWLEDGMENTS}
We thank Dr. Xianxin Wu and Prof. Xi Dai for helpful discussions.
This work was supported by the National Natural Science Foundation of China (11504117, 11774399, 11622435, U1832202), Beijing Natural Science Foundation (Z180008), the Ministry of Science and Technology of China (2016YFA0300600, 2016YFA0401000 and 2018YFA0305700), the Chinese Academy of Sciences (XDB28000000, XDB07000000), the Beijing Municipal Science and Technology Commission (Z181100004218001, Z171100002017018). H.W. acknowledges support from the Science Challenge Project (No.~TZ2016004), the K. C. Wong Education Foundation (GJTD-2018-01). Y.S. acknowledges the National Key Research and Development Program of China (No.~2017YFA0302901). Z.W. acknowledges support from the National Thousand-Young-Talents Program and the CAS Pioneer Hundred Talents Program.

\section*{AUTHOR CONTRIBUTIONS}
Z. W. and H. W. conceived and designed the project. Y. Q., S. N. and Z. W. performed all the DFT calculations,
Y. Q., S. N., C. F, Z. W., H. W. and Z. F. did the theoretical analysis.
C. Y. and Y. S. contributed in sample growth. L. K., T. Q. and H. D. carried out the ARPES experiment.
All authors contributed to the manuscript writing.

\section*{ADDITIONAL INFORMATION}
\noindent \textbf{Supplementary information} 
is available for this paper at \url{https://doi.org/10.1038/s41524-019-0260-6}. \\
\noindent \textbf{Competing Interests} the Authors declare no Competing Financial or Non-Financial Interests.


\begin{thebibliography}{10}
\expandafter\ifx\csname url\endcsname\relax
  \def\url#1{\texttt{#1}}\fi
\expandafter\ifx\csname urlprefix\endcsname\relax\def\urlprefix{URL }\fi
\providecommand{\bibinfo}[2]{#2}
\providecommand{\eprint}[2][]{\url{#2}}

\bibitem{kane2010}
\bibinfo{author}{Hasan, M.~Z.} \& \bibinfo{author}{Kane, C.~L.}
\newblock \bibinfo{title}{Colloquium: Topological insulators}.
\newblock \emph{\bibinfo{journal}{Rev. Mod. Phys.}}
  \textbf{\bibinfo{volume}{82}}, \bibinfo{pages}{3045--3067}
  (\bibinfo{year}{2010}).

\bibitem{qi2011}
\bibinfo{author}{Qi, X.-L.} \& \bibinfo{author}{Zhang, S.-C.}
\newblock \bibinfo{title}{Topological insulators and superconductors}.
\newblock \emph{\bibinfo{journal}{Rev. Mod. Phys.}}
  \textbf{\bibinfo{volume}{83}}, \bibinfo{pages}{1057--1110}
  (\bibinfo{year}{2011}).

\bibitem{wan2011topological}
\bibinfo{author}{Wan, X.}, \bibinfo{author}{Turner, A.~M.},
  \bibinfo{author}{Vishwanath, A.} \& \bibinfo{author}{Savrasov, S.~Y.}
\newblock \bibinfo{title}{Topological semimetal and Fermi-arc surface states in
  the electronic structure of pyrochlore iridates}.
\newblock \emph{\bibinfo{journal}{Phys. Rev. B}}
  \textbf{\bibinfo{volume}{83}}, \bibinfo{pages}{205101}
  (\bibinfo{year}{2011}).

\bibitem{xu2011chern}
\bibinfo{author}{Xu, G.}, \bibinfo{author}{Weng, H.}, \bibinfo{author}{Wang,
  Z.}, \bibinfo{author}{Dai, X.} \& \bibinfo{author}{Fang, Z.}
\newblock \bibinfo{title}{Chern semimetal and the quantized anomalous Hall
  effect in HgCr$_2$Se$_4$}.
\newblock \emph{\bibinfo{journal}{Phys. Rev. Lett.}}
  \textbf{\bibinfo{volume}{107}}, \bibinfo{pages}{186806}
  (\bibinfo{year}{2011}).

\bibitem{wang2012dirac}
\bibinfo{author}{Wang, Z.} \emph{et~al.}
\newblock \bibinfo{title}{Dirac semimetal and topological phase transitions in
  A$_3$Bi (A= Na, K, Rb)}.
\newblock \emph{\bibinfo{journal}{Phys. Rev. B}}
  \textbf{\bibinfo{volume}{85}}, \bibinfo{pages}{195320}
  (\bibinfo{year}{2012}).

\bibitem{wang2013three}
\bibinfo{author}{Wang, Z.}, \bibinfo{author}{Weng, H.}, \bibinfo{author}{Wu,
  Q.}, \bibinfo{author}{Dai, X.} \& \bibinfo{author}{Fang, Z.}
\newblock \bibinfo{title}{Three-dimensional Dirac semimetal and quantum
  transport in Cd$_3$As$_2$}.
\newblock \emph{\bibinfo{journal}{Phys. Rev. B}}
  \textbf{\bibinfo{volume}{88}}, \bibinfo{pages}{125427}
  (\bibinfo{year}{2013}).

\bibitem{weng2015weyl}
\bibinfo{author}{Weng, H.}, \bibinfo{author}{Fang, C.}, \bibinfo{author}{Fang,
  Z.}, \bibinfo{author}{Bernevig, B.~A.} \& \bibinfo{author}{Dai, X.}
\newblock \bibinfo{title}{Weyl semimetal phase in noncentrosymmetric
  transition-metal monophosphides}.
\newblock \emph{\bibinfo{journal}{Phys. Rev. X}}
  \textbf{\bibinfo{volume}{5}}, \bibinfo{pages}{011029} (\bibinfo{year}{2015}).

\bibitem{PhysRevLett.106.257004}
\bibinfo{author}{Pan, Z.-H.} \emph{et~al.}
\newblock \bibinfo{title}{Electronic structure of the topological insulator
  ${\mathrm{Bi}}_{2}{\mathrm{Se}}_{3}$ using angle-resolved photoemission
  spectroscopy: Evidence for a nearly full surface spin polarization}.
\newblock \emph{\bibinfo{journal}{Phys. Rev. Lett.}}
  \textbf{\bibinfo{volume}{106}}, \bibinfo{pages}{257004}
  (\bibinfo{year}{2011}).

\bibitem{zhang2013spin}
\bibinfo{author}{Zhang, H.}, \bibinfo{author}{Liu, C.-X.} \&
  \bibinfo{author}{Zhang, S.-C.}
\newblock \bibinfo{title}{Spin-orbital texture in topological insulators}.
\newblock \emph{\bibinfo{journal}{Phys. Rev. Lett.}}
  \textbf{\bibinfo{volume}{111}}, \bibinfo{pages}{066801}
  (\bibinfo{year}{2013}).

\bibitem{xu2016observation}
\bibinfo{author}{Xu, N.} \emph{et~al.}
\newblock \bibinfo{title}{Observation of Weyl nodes and Fermi arcs in tantalum
  phosphide}.
\newblock \emph{\bibinfo{journal}{Nat. Commun.}}
  \textbf{\bibinfo{volume}{7}}, \bibinfo{pages}{11006} (\bibinfo{year}{2016}).

\bibitem{xu2015discovery}
\bibinfo{author}{Xu, S.-Y.} \emph{et~al.}
\newblock \bibinfo{title}{Discovery of a Weyl fermion semimetal and topological
  Fermi arcs}.
\newblock \emph{\bibinfo{journal}{Science}} \textbf{\bibinfo{volume}{349}},
  \bibinfo{pages}{613--617} (\bibinfo{year}{2015}).

\bibitem{wang2016observation2}
\bibinfo{author}{Wang, C.} \emph{et~al.}
\newblock \bibinfo{title}{Observation of Fermi arc and its connection with bulk
  states in the candidate type-II Weyl semimetal WTe$_2$}.
\newblock \emph{\bibinfo{journal}{Phys. Rev. B}}
  \textbf{\bibinfo{volume}{94}}, \bibinfo{pages}{241119}
  (\bibinfo{year}{2016}).

\bibitem{huang2015observation}
\bibinfo{author}{Huang, X.} \emph{et~al.}
\newblock \bibinfo{title}{Observation of the chiral-anomaly-induced negative
  magnetoresistance in 3D Weyl semimetal TaAs}.
\newblock \emph{\bibinfo{journal}{Phys. Rev. X}}
  \textbf{\bibinfo{volume}{5}}, \bibinfo{pages}{031023} (\bibinfo{year}{2015}).

\bibitem{zhang2016signatures}
\bibinfo{author}{Zhang, C.-L.} \emph{et~al.}
\newblock \bibinfo{title}{Signatures of the Adler--Bell--Jackiw chiral anomaly
  in a Weyl fermion semimetal}.
\newblock \emph{\bibinfo{journal}{Nat. Commun.}}
  \textbf{\bibinfo{volume}{7}}, \bibinfo{pages}{10735} (\bibinfo{year}{2016}).

\bibitem{PhysRevLett.98.106803}
\bibinfo{author}{Fu, L.}, \bibinfo{author}{Kane, C.~L.} \&
  \bibinfo{author}{Mele, E.~J.}
\newblock \bibinfo{title}{Topological insulators in three dimensions}.
\newblock \emph{\bibinfo{journal}{Phys. Rev. Lett.}}
  \textbf{\bibinfo{volume}{98}}, \bibinfo{pages}{106803}
  (\bibinfo{year}{2007}).

\bibitem{hsieh2012topological}
\bibinfo{author}{Hsieh, T.~H.} \emph{et~al.}
\newblock \bibinfo{title}{Topological crystalline insulators in the SnTe
  material class}.
\newblock \emph{\bibinfo{journal}{Nat. Commun.}}
  \textbf{\bibinfo{volume}{3}}, \bibinfo{pages}{982} (\bibinfo{year}{2012}).

\bibitem{nie2016band}
\bibinfo{author}{Nie, S.}, \bibinfo{author}{Xu, X.~Y.}, \bibinfo{author}{Xu,
  G.} \& \bibinfo{author}{Fang, Z.}
\newblock \bibinfo{title}{Band gap anomaly and topological properties in lead
  chalcogenides}.
\newblock \emph{\bibinfo{journal}{Chinese Phys. B}}
  \textbf{\bibinfo{volume}{25}}, \bibinfo{pages}{037311}
  (\bibinfo{year}{2016}).

\bibitem{soluyanov2015type}
\bibinfo{author}{Soluyanov, A.~A.} \emph{et~al.}
\newblock \bibinfo{title}{Type-II Weyl semimetals}.
\newblock \emph{\bibinfo{journal}{Nature}} \textbf{\bibinfo{volume}{527}},
  \bibinfo{pages}{495} (\bibinfo{year}{2015}).

\bibitem{huang2015weyl}
\bibinfo{author}{Huang, S.-M.} \emph{et~al.}
\newblock \bibinfo{title}{A Weyl fermion semimetal with surface Fermi arcs in
  the transition metal monopnictide TaAs class}.
\newblock \emph{\bibinfo{journal}{Nat. Commun.}}
  \textbf{\bibinfo{volume}{6}}, \bibinfo{pages}{7373} (\bibinfo{year}{2015}).

\bibitem{lv2015experimental}
\bibinfo{author}{Lv, B.} \emph{et~al.}
\newblock \bibinfo{title}{Experimental discovery of Weyl semimetal TaAs}.
\newblock \emph{\bibinfo{journal}{Phys. Rev. X}}
  \textbf{\bibinfo{volume}{5}}, \bibinfo{pages}{031013} (\bibinfo{year}{2015}).

\bibitem{lv2015observation}
\bibinfo{author}{Lv, B.} \emph{et~al.}
\newblock \bibinfo{title}{Observation of Weyl nodes in TaAs}.
\newblock \emph{\bibinfo{journal}{Nat. Phys.}}
  \textbf{\bibinfo{volume}{11}}, \bibinfo{pages}{724} (\bibinfo{year}{2015}).

\bibitem{lv2015observation2}
\bibinfo{author}{Lv, B.} \emph{et~al.}
\newblock \bibinfo{title}{Observation of Fermi-arc spin texture in TaAs}.
\newblock \emph{\bibinfo{journal}{Phys. Rev. Lett.}}
  \textbf{\bibinfo{volume}{115}}, \bibinfo{pages}{217601}
  (\bibinfo{year}{2015}).

\bibitem{nie2017topological}
\bibinfo{author}{Nie, S.}, \bibinfo{author}{Xu, G.}, \bibinfo{author}{Prinz,
  F.~B.} \& \bibinfo{author}{Zhang, S.-C.}
\newblock \bibinfo{title}{Topological semimetal in honeycomb lattice LnSI}.
\newblock \emph{\bibinfo{journal}{Proc. Natl Acad. Sci. USA}} \textbf{\bibinfo{volume}{114}}, \bibinfo{pages}{10596--10600}
  (\bibinfo{year}{2017}).

\bibitem{nie2019magnetic}
\bibinfo{author}{Nie, S.} \emph{et~al.}
\newblock \bibinfo{title}{Magnetic semimetals and quantized anomalous Hall
  effect in EuB$_6$}.
\newblock \emph{\bibinfo{journal}{arXiv preprint arXiv:1907.10051}}
  (\bibinfo{year}{2019}).

\bibitem{PhysRevLett.117.236401}
\bibinfo{author}{Wang, Z.} \emph{et~al.}
\newblock \bibinfo{title}{Time-reversal-breaking Weyl fermions in magnetic
  Heusler alloys}.
\newblock \emph{\bibinfo{journal}{Phys. Rev. Lett.}}
  \textbf{\bibinfo{volume}{117}}, \bibinfo{pages}{236401}
  (\bibinfo{year}{2016}).

\bibitem{deng2016experimental}
\bibinfo{author}{Deng, K.} \emph{et~al.}
\newblock \bibinfo{title}{Experimental observation of topological Fermi arcs in
  type-II Weyl semimetal MoTe$_2$}.
\newblock \emph{\bibinfo{journal}{Nat. Phys.}}
  \textbf{\bibinfo{volume}{12}}, \bibinfo{pages}{1105} (\bibinfo{year}{2016}).

\bibitem{jiang2017signature}
\bibinfo{author}{Jiang, J.} \emph{et~al.}
\newblock \bibinfo{title}{Signature of type-II Weyl semimetal phase in MoTe$_2$}.
\newblock \emph{\bibinfo{journal}{Nat. Commun.}}
  \textbf{\bibinfo{volume}{8}}, \bibinfo{pages}{13973} (\bibinfo{year}{2017}).

\bibitem{tamai2016fermi}
\bibinfo{author}{Tamai, A.} \emph{et~al.}
\newblock \bibinfo{title}{Fermi arcs and their topological character in the
  candidate type-II Weyl semimetal MoTe$_2$}.
\newblock \emph{\bibinfo{journal}{Phys. Rev. X}}
  \textbf{\bibinfo{volume}{6}}, \bibinfo{pages}{031021} (\bibinfo{year}{2016}).

\bibitem{liang2016electronic}
\bibinfo{author}{Liang, A.} \emph{et~al.}
\newblock \bibinfo{title}{Electronic evidence for type II Weyl semimetal state
  in MoTe$_2$}.
\newblock \emph{\bibinfo{journal}{arXiv preprint arXiv:1604.01706}}
  (\bibinfo{year}{2016}).

\bibitem{wang2016mote}
\bibinfo{author}{Wang, Z.} \emph{et~al.}
\newblock \bibinfo{title}{Mote 2: a type-II Weyl topological metal}.
\newblock \emph{\bibinfo{journal}{Phys. Rev. Lett.}}
  \textbf{\bibinfo{volume}{117}}, \bibinfo{pages}{056805}
  (\bibinfo{year}{2016}).

\bibitem{kumar2017extremely}
\bibinfo{author}{Kumar, N.} \emph{et~al.}
\newblock \bibinfo{title}{Extremely high magnetoresistance and conductivity in
  the type-II Weyl semimetals WP$_2$ and MoP$_2$}.
\newblock \emph{\bibinfo{journal}{Nat. Commun.}}
  \textbf{\bibinfo{volume}{8}}, \bibinfo{pages}{1642} (\bibinfo{year}{2017}).

\bibitem{shekhar2015extremely}
\bibinfo{author}{Shekhar, C.} \emph{et~al.}
\newblock \bibinfo{title}{Extremely large magnetoresistance and ultrahigh
  mobility in the topological Weyl semimetal candidate NbP}.
\newblock \emph{\bibinfo{journal}{Nat. Phys.}}
  \textbf{\bibinfo{volume}{11}}, \bibinfo{pages}{645} (\bibinfo{year}{2015}).

\bibitem{yan2013large}
\bibinfo{author}{Yan, B.}, \bibinfo{author}{Jansen, M.} \&
  \bibinfo{author}{Felser, C.}
\newblock \bibinfo{title}{A large-energy-gap oxide topological insulator based
  on the superconductor BaBiO$_3$}.
\newblock \emph{\bibinfo{journal}{Nat. Phys.}}
  \textbf{\bibinfo{volume}{9}}, \bibinfo{pages}{709} (\bibinfo{year}{2013}).

\bibitem{wu2015}
\bibinfo{author}{Wu, X.} \emph{et~al.}
\newblock \bibinfo{title}{${\mathrm{cafeas}}_{2}$: A staggered intercalation of
  quantum spin Hall and high-temperature superconductivity}.
\newblock \emph{\bibinfo{journal}{Phys. Rev. B}} \textbf{\bibinfo{volume}{91}},
  \bibinfo{pages}{081111} (\bibinfo{year}{2015}).

\bibitem{Schoop2015}
\bibinfo{author}{Schoop, L.~M.} \emph{et~al.}
\newblock \bibinfo{title}{Dirac metal to topological metal transition at a
  structural phase change in ${\mathrm{Au}}_{2}\mathrm{Pb}$ and prediction of
  ${\mathbb{Z}}_{2}$ topology for the superconductor}.
\newblock \emph{\bibinfo{journal}{Phys. Rev. B}} \textbf{\bibinfo{volume}{91}},
  \bibinfo{pages}{214517} (\bibinfo{year}{2015}).

\bibitem{chang2016}
\bibinfo{author}{Chang, T.-R.} \emph{et~al.}
\newblock \bibinfo{title}{Topological Dirac surface states and superconducting
  pairing correlations in ${\mathrm{PbTaSe}}_{2}$}.
\newblock \emph{\bibinfo{journal}{Phys. Rev. B}} \textbf{\bibinfo{volume}{93}},
  \bibinfo{pages}{245130} (\bibinfo{year}{2016}).

\bibitem{wang2016spontaneous}
\bibinfo{author}{Wang, Y.-Q.} \emph{et~al.}
\newblock \bibinfo{title}{Spontaneous formation of a
  superconductor--topological insulator--normal metal layered heterostructure}.
\newblock \emph{\bibinfo{journal}{Adv. Mater.}}
  \textbf{\bibinfo{volume}{28}}, \bibinfo{pages}{5013--5017}
  (\bibinfo{year}{2016}).

\bibitem{xie2017}
\bibinfo{author}{Bian, G.} \emph{et~al.}
\newblock \bibinfo{title}{Prediction of nontrivial band topology and
  superconductivity in $\mathrm{M}{\mathrm{g}}_{2}\mathrm{Pb}$}.
\newblock \emph{\bibinfo{journal}{Phys. Rev. Mater.}}
  \textbf{\bibinfo{volume}{1}}, \bibinfo{pages}{021201} (\bibinfo{year}{2017}).

\bibitem{nie2018}
\bibinfo{author}{Nie, S.} \emph{et~al.}
\newblock \bibinfo{title}{Topological phases in the ${\mathrm{TaSe}}_{3}$
  compound}.
\newblock \emph{\bibinfo{journal}{Phys. Rev. B}} \textbf{\bibinfo{volume}{98}},
  \bibinfo{pages}{125143} (\bibinfo{year}{2018}).

\bibitem{Fu2008superconducting}
\bibinfo{author}{Fu, L.} \& \bibinfo{author}{Kane, C.~L.}
\newblock \bibinfo{title}{Superconducting proximity effect and Majorana
  fermions at the surface of a topological insulator}.
\newblock \emph{\bibinfo{journal}{Phys. Rev. Lett.}}
  \textbf{\bibinfo{volume}{100}}, \bibinfo{pages}{096407}
  (\bibinfo{year}{2008}).

\bibitem{Fu2010odd}
\bibinfo{author}{Fu, L.} \& \bibinfo{author}{Berg, E.}
\newblock \bibinfo{title}{Odd-parity topological superconductors: Theory and
  application to ${\mathrm{Cu}}_{x}{\mathrm{Bi}}_{2}{\mathrm{Se}}_{3}$}.
\newblock \emph{\bibinfo{journal}{Phys. Rev. Lett.}}
  \textbf{\bibinfo{volume}{105}}, \bibinfo{pages}{097001}
  (\bibinfo{year}{2010}).

\bibitem{alicea2012new}
\bibinfo{author}{Alicea, J.}
\newblock \bibinfo{title}{New directions in the pursuit of Majorana fermions in
  solid state systems}.
\newblock \emph{\bibinfo{journal}{Rep. Prog. Phys.}}
  \textbf{\bibinfo{volume}{75}}, \bibinfo{pages}{076501}
  (\bibinfo{year}{2012}).

\bibitem{sato2017topological}
\bibinfo{author}{Sato, M.} \& \bibinfo{author}{Ando, Y.}
\newblock \bibinfo{title}{Topological superconductors: a review}.
\newblock \emph{\bibinfo{journal}{Rep. Prog. Phys.}}
  \textbf{\bibinfo{volume}{80}}, \bibinfo{pages}{076501}
  (\bibinfo{year}{2017}).

\bibitem{wang2015}
\bibinfo{author}{Wang, Z.} \emph{et~al.}
\newblock \bibinfo{title}{Topological nature of the
  ${\mathrm{FeSe}}_{0.5}{\mathrm{Te}}_{0.5}$ superconductor}.
\newblock \emph{\bibinfo{journal}{Phys. Rev. B}} \textbf{\bibinfo{volume}{92}},
  \bibinfo{pages}{115119} (\bibinfo{year}{2015}).

\bibitem{xu2016}
\bibinfo{author}{Xu, G.}, \bibinfo{author}{Lian, B.}, \bibinfo{author}{Tang,
  P.}, \bibinfo{author}{Qi, X.-L.} \& \bibinfo{author}{Zhang, S.-C.}
\newblock \bibinfo{title}{Topological superconductivity on the surface of
  Fe-based superconductors}.
\newblock \emph{\bibinfo{journal}{Phys. Rev. Lett.}}
  \textbf{\bibinfo{volume}{117}}, \bibinfo{pages}{047001}
  (\bibinfo{year}{2016}).

\bibitem{Zhang182}
\bibinfo{author}{Zhang, P.} \emph{et~al.}
\newblock \bibinfo{title}{Observation of topological superconductivity on the
  surface of an iron-based superconductor}.
\newblock \emph{\bibinfo{journal}{Science}} \textbf{\bibinfo{volume}{360}},
  \bibinfo{pages}{182--186} (\bibinfo{year}{2018}).

\bibitem{Wang333}
\bibinfo{author}{Wang, D.} \emph{et~al.}
\newblock \bibinfo{title}{Evidence for Majorana bound states in an iron-based
  superconductor}.
\newblock \emph{\bibinfo{journal}{Science}} \textbf{\bibinfo{volume}{362}},
  \bibinfo{pages}{333--335} (\bibinfo{year}{2018}).

\bibitem{qi2010}
\bibinfo{author}{Qi, X.-L.}, \bibinfo{author}{Hughes, T.~L.} \&
  \bibinfo{author}{Zhang, S.-C.}
\newblock \bibinfo{title}{Topological invariants for the Fermi surface of a
  time-reversal-invariant superconductor}.
\newblock \emph{\bibinfo{journal}{Phys. Rev. B}} \textbf{\bibinfo{volume}{81}},
  \bibinfo{pages}{134508} (\bibinfo{year}{2010}).

\bibitem{Hosur204}
\bibinfo{author}{Hosur, P.}, \bibinfo{author}{Dai, X.}, \bibinfo{author}{Fang,
  Z.} \& \bibinfo{author}{Qi, X.-L.}
\newblock \bibinfo{title}{Time-reversal-invariant topological superconductivity
  in doped Weyl semimetals}.
\newblock \emph{\bibinfo{journal}{Phys. Rev. B}} \textbf{\bibinfo{volume}{90}},
  \bibinfo{pages}{045130} (\bibinfo{year}{2014}).

\bibitem{pan2015pressure}
\bibinfo{author}{Pan, X.-C.} \emph{et~al.}
\newblock \bibinfo{title}{Pressure-driven dome-shaped superconductivity and
  electronic structural evolution in tungsten ditelluride}.
\newblock \emph{\bibinfo{journal}{Nat. Commun.}}
  \textbf{\bibinfo{volume}{6}}, \bibinfo{pages}{7805} (\bibinfo{year}{2015}).

\bibitem{kang2015superconductivity}
\bibinfo{author}{Kang, D.} \emph{et~al.}
\newblock \bibinfo{title}{Superconductivity emerging from a suppressed large
  magnetoresistant state in tungsten ditelluride}.
\newblock \emph{\bibinfo{journal}{Nat. Commun.}}
  \textbf{\bibinfo{volume}{6}}, \bibinfo{pages}{7804} (\bibinfo{year}{2015}).

\bibitem{qi2016superconductivity}
\bibinfo{author}{Qi, Y.} \emph{et~al.}
\newblock \bibinfo{title}{Superconductivity in Weyl semimetal candidate MoTe$_2$}.
\newblock \emph{\bibinfo{journal}{Nat. Commun.}}
  \textbf{\bibinfo{volume}{7}}, \bibinfo{pages}{11038} (\bibinfo{year}{2016}).

\bibitem{chen2016superconductivity}
\bibinfo{author}{Chen, F.} \emph{et~al.}
\newblock \bibinfo{title}{Superconductivity enhancement in the s-doped Weyl
  semimetal candidate MoTe$_2$}.
\newblock \emph{\bibinfo{journal}{Appl. Phys. Lett.}}
  \textbf{\bibinfo{volume}{108}}, \bibinfo{pages}{162601}
  (\bibinfo{year}{2016}).

\bibitem{li2017concurrence}
\bibinfo{author}{Li, Y.} \emph{et~al.}
\newblock \bibinfo{title}{Concurrence of superconductivity and structure
  transition in Weyl semimetal TaP under pressure}.
\newblock \emph{\bibinfo{journal}{npj Quantum Mater.}}
  \textbf{\bibinfo{volume}{2}}, \bibinfo{pages}{66} (\bibinfo{year}{2017}).

\bibitem{xu2019topological}
\bibinfo{author}{Xu, Y.} \emph{et~al.}
\newblock \bibinfo{title}{Topological nodal lines and hybrid Weyl nodes in YCoC$_2$}.
\newblock \emph{\bibinfo{journal}{APL Mater.}}
\textbf{\bibinfo{volume}{7}}, \bibinfo{pages}{101109} (\bibinfo{year}{2019}).

\bibitem{barz1980ternary}
\bibinfo{author}{Barz, H.}, \bibinfo{author}{Ku, H.}, \bibinfo{author}{Meisner,
  G.}, \bibinfo{author}{Fisk, Z.} \& \bibinfo{author}{Matthias, B.}
\newblock \bibinfo{title}{Ternary transition metal phosphides: High-temperature
  superconductors}.
\newblock \emph{\bibinfo{journal}{Proc. Natl Acad. Sci. USA}} \textbf{\bibinfo{volume}{77}}, \bibinfo{pages}{3132--3134}
  (\bibinfo{year}{1980}).

\bibitem{meisner1983superconductivity}
\bibinfo{author}{Meisner, G.} \& \bibinfo{author}{Ku, H.}
\newblock \bibinfo{title}{The superconductivity and structure of equiatomic
  ternary transition metal pnictides}.
\newblock \emph{\bibinfo{journal}{Appl. Phys. A}}
  \textbf{\bibinfo{volume}{31}}, \bibinfo{pages}{201--212}
  (\bibinfo{year}{1983}).

\bibitem{MULLER1983177}
\bibinfo{author}{M$\ddot{\text{u}}$ller, R.}, \bibinfo{author}{Shelton, R.},
  \bibinfo{author}{Richardson, J.} \& \bibinfo{author}{Jacobson, R.}
\newblock \bibinfo{title}{Superconductivity and crystal structure of a new
  class of ternary transition metal phosphides TT'P (T=Zr, Nb, Ta and T'=Ru,
  Rh)}.
\newblock \emph{\bibinfo{journal}{J. Less Common Metals}}
  \textbf{\bibinfo{volume}{92}}, \bibinfo{pages}{177 -- 183}
  (\bibinfo{year}{1983}).

\bibitem{Fu2007topo}
\bibinfo{author}{Fu, L.}, \bibinfo{author}{Kane, C.~L.} \&
  \bibinfo{author}{Mele, E.~J.}
\newblock \bibinfo{title}{Topological insulators in three dimensions}.
\newblock \emph{\bibinfo{journal}{Phys. Rev. Lett.}}
  \textbf{\bibinfo{volume}{98}}, \bibinfo{pages}{106803}
  (\bibinfo{year}{2007}).

\bibitem{yamakage2015line}
\bibinfo{author}{Yamakage, A.}, \bibinfo{author}{Yamakawa, Y.},
  \bibinfo{author}{Tanaka, Y.} \& \bibinfo{author}{Okamoto, Y.}
\newblock \bibinfo{title}{Line-node Dirac semimetal and topological insulating
  phase in noncentrosymmetric pnictides CaAgX (X=P, As)}.
\newblock \emph{\bibinfo{journal}{J. Phys. Soc. Jpn.}}
  \textbf{\bibinfo{volume}{85}}, \bibinfo{pages}{013708}
  (\bibinfo{year}{2015}).

\bibitem{tqc2017}
\bibinfo{author}{Bradlyn, B.} \emph{et~al.}
\newblock \bibinfo{title}{Topological quantum chemistry}.
\newblock \emph{\bibinfo{journal}{Nature}} \textbf{\bibinfo{volume}{547}},
  \bibinfo{pages}{298--305} (\bibinfo{year}{2017}).

\bibitem{wang2019}
\bibinfo{author}{Vergniory, M.} \emph{et~al.}
\newblock \bibinfo{title}{A complete catalogue of high-quality topological
  materials}.
\newblock \emph{\bibinfo{journal}{Nature}} \textbf{\bibinfo{volume}{566}},
  \bibinfo{pages}{480--485} (\bibinfo{year}{2019}).

\bibitem{ashvin2017}
\bibinfo{author}{Po, H.~C.}, \bibinfo{author}{Vishwanath, A.} \&
  \bibinfo{author}{Watanabe, H.}
\newblock \bibinfo{title}{Complete theory of symmetry-based indicators of band
  topology}.
\newblock \emph{\bibinfo{journal}{Nat. Commun.}}
  \textbf{\bibinfo{volume}{8}}, \bibinfo{pages}{50} (\bibinfo{year}{2017}).

\bibitem{song2017}
\bibinfo{author}{{Song}, Z.}, \bibinfo{author}{{Zhang}, T.},
  \bibinfo{author}{{Fang}, Z.} \& \bibinfo{author}{{Fang}, C.}
\newblock \bibinfo{title}{{Quantitative mappings between symmetry and topology
  in solids }}.
\newblock \emph{\bibinfo{journal}{Nat. Commun.}}
  \textbf{\bibinfo{volume}{9}}, \bibinfo{pages}{3530} (\bibinfo{year}{2018}).

\bibitem{Jorrit2017}
\bibinfo{author}{Kruthoff, J.}, \bibinfo{author}{de~Boer, J.},
  \bibinfo{author}{van Wezel, J.}, \bibinfo{author}{Kane, C.~L.} \&
  \bibinfo{author}{Slager, R.-J.}
\newblock \bibinfo{title}{Topological classification of crystalline insulators
  through band structure combinatorics}.
\newblock \emph{\bibinfo{journal}{Phys. Rev. X}} \textbf{\bibinfo{volume}{7}},
  \bibinfo{pages}{041069} (\bibinfo{year}{2017}).

\bibitem{zhang2019}
\bibinfo{author}{{Zhang}, T.} \emph{et~al.}
\newblock \bibinfo{title}{Catalogue of topological electronic materials}.
\newblock \emph{\bibinfo{journal}{Nature}} \textbf{\bibinfo{volume}{566}},
  \bibinfo{pages}{475} (\bibinfo{year}{2019}).

\bibitem{wanxg2019}
\bibinfo{author}{Tang, F.}, \bibinfo{author}{Po, H.~C.},
  \bibinfo{author}{Vishwanath, A.} \& \bibinfo{author}{Wan, X.}
\newblock \bibinfo{title}{{Comprehensive search for topological materials using
  symmetry indicators}}.
\newblock \emph{\bibinfo{journal}{Nature}} \textbf{\bibinfo{volume}{566}},
  \bibinfo{pages}{486--489} (\bibinfo{year}{2019}).


\bibitem{bernevig2015s}
\bibinfo{author}{Bernevig, B.~A.}
\newblock \bibinfo{title}{It's been a Weyl coming}.
\newblock \emph{\bibinfo{journal}{Nat. Phys.}}
  \textbf{\bibinfo{volume}{11}}, \bibinfo{pages}{698} (\bibinfo{year}{2015}).

\bibitem{xu2015}
\bibinfo{author}{Xu, Y.}, \bibinfo{author}{Zhang, F.} \&
  \bibinfo{author}{Zhang, C.}
\newblock \bibinfo{title}{Structured Weyl points in spin-orbit coupled
  fermionic superfluids}.
\newblock \emph{\bibinfo{journal}{Phys. Rev. Lett.}}
  \textbf{\bibinfo{volume}{115}}, \bibinfo{pages}{265304}
  (\bibinfo{year}{2015}).

\bibitem{Sancho_1985}
\bibinfo{author}{Sancho, M. P.~L.}, \bibinfo{author}{Sancho, J. M.~L.},
  \bibinfo{author}{Sancho, J. M.~L.} \& \bibinfo{author}{Rubio, J.}
\newblock \bibinfo{title}{Highly convergent schemes for the calculation of bulk
  and surface green functions}.
\newblock \emph{\bibinfo{journal}{J. Phys. F: Metal Phys.}}
  \textbf{\bibinfo{volume}{15}}, \bibinfo{pages}{851--858}
  (\bibinfo{year}{1985}).

\bibitem{shingo2015}
\bibinfo{author}{Kobayashi, S.} \& \bibinfo{author}{Sato, M.}
\newblock \bibinfo{title}{Topological superconductivity in Dirac semimetals}.
\newblock \emph{\bibinfo{journal}{Phys. Rev. Lett.}}
  \textbf{\bibinfo{volume}{115}}, \bibinfo{pages}{187001}
  (\bibinfo{year}{2015}).

\bibitem{ueno2013}
\bibinfo{author}{Ueno, Y.}, \bibinfo{author}{Yamakage, A.},
  \bibinfo{author}{Tanaka, Y.} \& \bibinfo{author}{Sato, M.}
\newblock \bibinfo{title}{Symmetry-protected Majorana fermions in topological
  crystalline superconductors: Theory and application to
  ${\mathrm{Sr}}_{2}{\mathrm{RuO}}_{4}$}.
\newblock \emph{\bibinfo{journal}{Phys. Rev. Lett.}}
  \textbf{\bibinfo{volume}{111}}, \bibinfo{pages}{087002}
  (\bibinfo{year}{2013}).

\bibitem{ivanov2019}
\bibinfo{author}{Ivanov, V.} \& \bibinfo{author}{Savrasov, S.~Y.}
\newblock \bibinfo{title}{Monopole mining method for high-throughput screening
  for Weyl semimetals}.
\newblock \emph{\bibinfo{journal}{Phys. Rev. B}} \textbf{\bibinfo{volume}{99}},
  \bibinfo{pages}{125124} (\bibinfo{year}{2019}).

\bibitem{paw1}
\bibinfo{author}{Bl\"ochl, P.~E.}
\newblock \bibinfo{title}{Projector augmented-wave method}.
\newblock \emph{\bibinfo{journal}{Phys. Rev. B}} \textbf{\bibinfo{volume}{50}},
  \bibinfo{pages}{17953--17979} (\bibinfo{year}{1994}).

\bibitem{paw2}
\bibinfo{author}{Kresse, G.} \& \bibinfo{author}{Joubert, D.}
\newblock \bibinfo{title}{From ultrasoft pseudopotentials to the projector
  augmented-wave method}.
\newblock \emph{\bibinfo{journal}{Phys. Rev. B}} \textbf{\bibinfo{volume}{59}},
  \bibinfo{pages}{1758--1775} (\bibinfo{year}{1999}).

\bibitem{KRESSE199615}
\bibinfo{author}{Kresse, G.} \& \bibinfo{author}{Furthmüller, J.}
\newblock \bibinfo{title}{Efficiency of ab-initio total energy calculations for
  metals and semiconductors using a plane-wave basis set}.
\newblock \emph{\bibinfo{journal}{Comput. Mater. Sci.}}
  \textbf{\bibinfo{volume}{6}}, \bibinfo{pages}{15 -- 50}
  (\bibinfo{year}{1996}).

\bibitem{vasp}
\bibinfo{author}{Kresse, G.} \& \bibinfo{author}{Furthm\"uller, J.}
\newblock \bibinfo{title}{Efficient iterative schemes for {\it{ab initio}}
  total-energy calculations using a plane-wave basis set}.
\newblock \emph{\bibinfo{journal}{Phys. Rev. B}} \textbf{\bibinfo{volume}{54}},
  \bibinfo{pages}{11169--11186} (\bibinfo{year}{1996}).

\bibitem{pbe}
\bibinfo{author}{Perdew, J.~P.}, \bibinfo{author}{Burke, K.} \&
  \bibinfo{author}{Ernzerhof, M.}
\newblock \bibinfo{title}{Generalized gradient approximation made simple}.
\newblock \emph{\bibinfo{journal}{Phys. Rev. Lett.}}
  \textbf{\bibinfo{volume}{77}}, \bibinfo{pages}{3865--3868}
  (\bibinfo{year}{1996}).

\bibitem{Meisner1983}
\bibinfo{author}{Meisner, G.~P.} \& \bibinfo{author}{Ku, H.~C.}
\newblock \bibinfo{title}{The superconductivity and structure of equiatomic
  ternary transition metal pnictides}.
\newblock \emph{\bibinfo{journal}{Appl. Phys. A}}
  \textbf{\bibinfo{volume}{31}}, \bibinfo{pages}{201--212}
  (\bibinfo{year}{1983}).

\bibitem{MEISNER1983983}
\bibinfo{author}{Meisner, G.}, \bibinfo{author}{Ku, H.} \&
  \bibinfo{author}{Barz, H.}
\newblock \bibinfo{title}{Superconducting equiatomic ternary transition metal
  arsenides}.
\newblock \emph{\bibinfo{journal}{Mater. Res. Bull.}}
  \textbf{\bibinfo{volume}{18}}, \bibinfo{pages}{983 -- 991}
  (\bibinfo{year}{1983}).

\bibitem{mlwf}
\bibinfo{author}{Marzari, N.}, \bibinfo{author}{Mostofi, A.~A.},
  \bibinfo{author}{Yates, J.~R.}, \bibinfo{author}{Souza, I.} \&
  \bibinfo{author}{Vanderbilt, D.}
\newblock \bibinfo{title}{Maximally localized Wannier functions: Theory and
  applications}.
\newblock \emph{\bibinfo{journal}{Rev. Mod. Phys.}}
  \textbf{\bibinfo{volume}{84}}, \bibinfo{pages}{1419--1475}
  (\bibinfo{year}{2012}).

\end{thebibliography}

\vspace{5mm}

\begin{figure}[!h]
\centering
\includegraphics[width=16 cm]{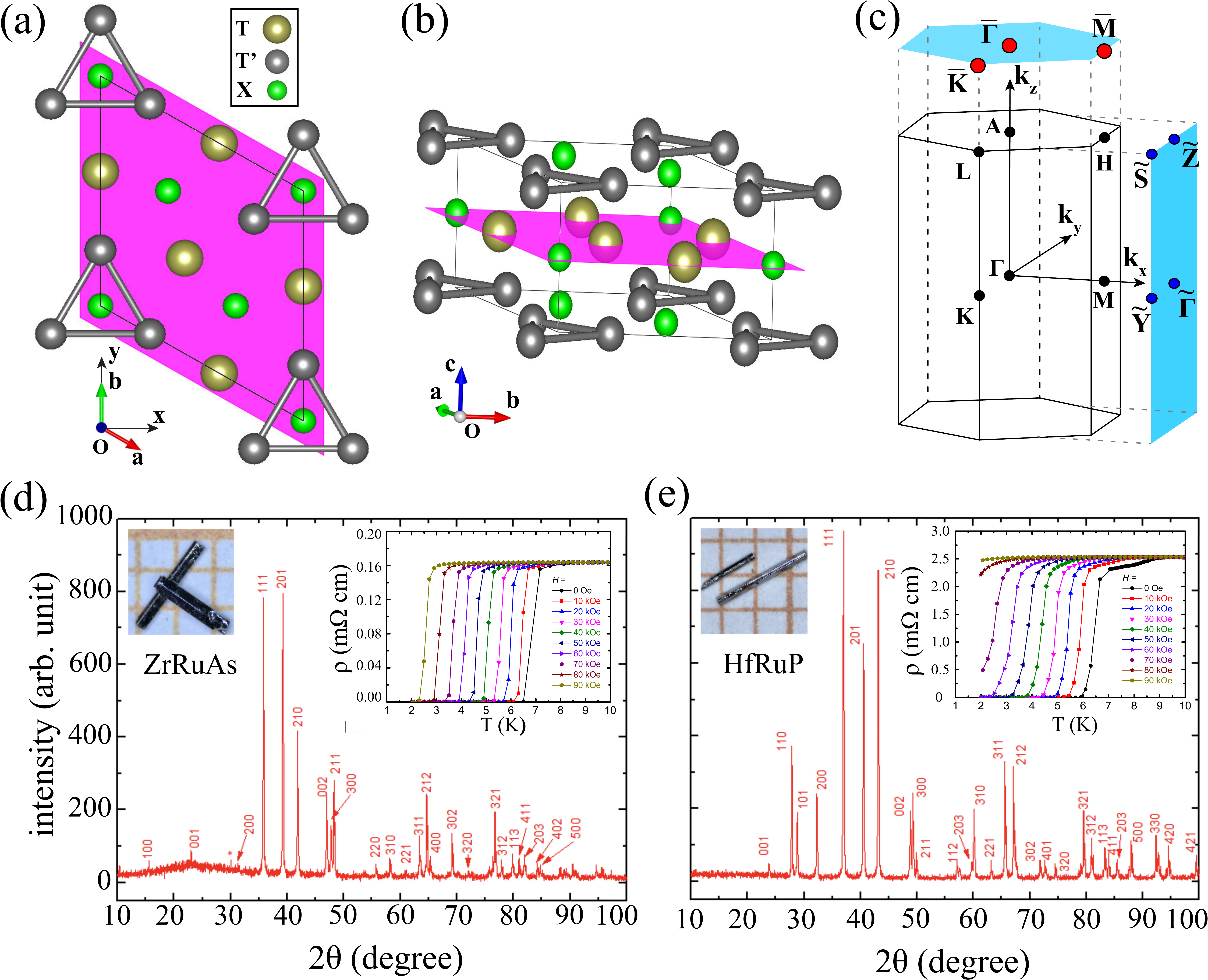}
\caption{(Color online)
The crystal structure, BZs, XRD spectra and transport properties of TT'X.
($\bold a$) The top view of the crystal, with brown, gray and green balls representing T, T' and X atoms, respectively.
($\bold b$) The perspective view of the crystal. 
($\bold c$) The bulk BZ, (001)-surface BZ and (100)-surface BZ. Hereafter, (001) [(100) or (010)] refers to the surface normal vector in terms of the Cartesian coordinates.
($\bold d$) and ($\bold e$) Indexed powder XRD spectra of ZrRuAs and HfRuP, respectively. Red stars are small amount of impurities. The left insets of ($\bold d$) and ($\bold e$) are photographs of ZrRuAs and HfRuP single crystals, respectively.  The right insets of them are temperature dependent longitudinal resistivity of ZrRuAs and HfRuP at various magnetic fields. The magnetic fields are perpendicular to $c$ axis and electronic current direction.
}
\end{figure}

\begin{figure}[!h]
\centering
\includegraphics[width=16 cm]{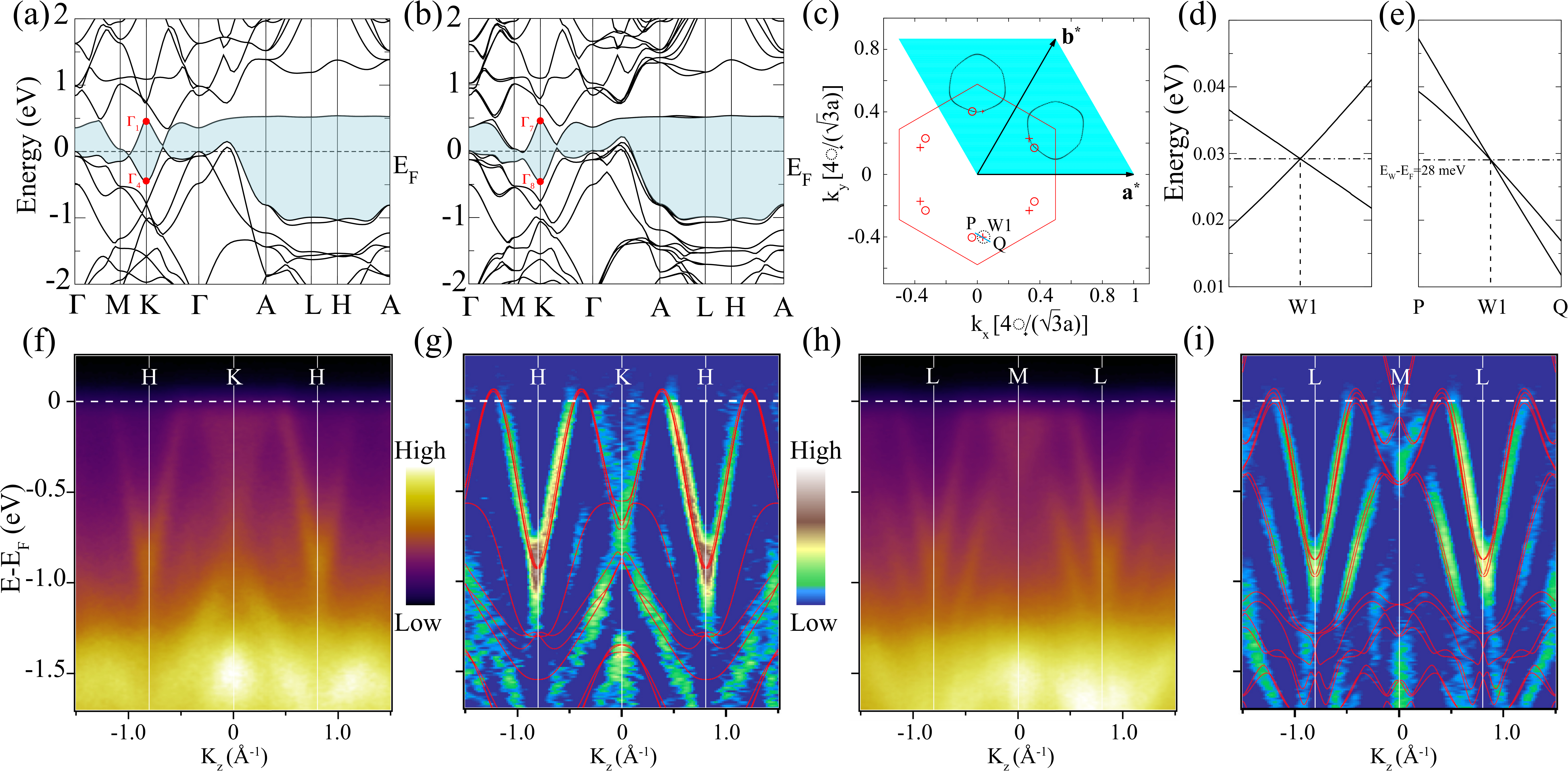}
\caption{(Color online)
The electronic band structures of HfRuP without ($\bold a$) and with ($\bold b$) SOC.
The irreducible representations of selected bands at K point are indicated.
($\bold c$) The nodal lines are presented in the $k_z=0$ plane without SOC.
With SOC, the WPs above the $k_z=0$ plane are labeled as ``+"(+1) and ``o"(-1) in the first BZ.
Here, $\bf a$*, $\bf b$*, and $\bf c$* are the reciprocal primitive vectors.
($\bold d$) The $k_z$ dispersion of the electronic bands through the WP W1 shown in ($\bold c$).
($\bold e$) The dispersion of the WP W1 along the line P-Q shown in ($\bold c$).
The fractional coordinates of W1, P and Q are $(0.2761 , -0.4654, 0.02439)$, $(0.2603 , -0.4603, 0.02439)$, and $(0.2919 , -0.4705, 0.02439)$. Hereafter, the positions of $k$-points are given in units of ($\bf a$*, $\bf b$*, $\bf c$*).
ARPES spectrum ($\bold f$) and curvature intensity ($\bold g$) plots of ZrRuAs, showing band structure along H-K-H. For comparison, the calculated band structure along H-K-H is superposed on the experimental data in ($\bold g$). ($\bold {h-i}$) are the same as ($\bold {f-g}$), but along L-M-L.
}
\end{figure}

\begin{figure}[!h]
\centering
\includegraphics[width=16 cm]{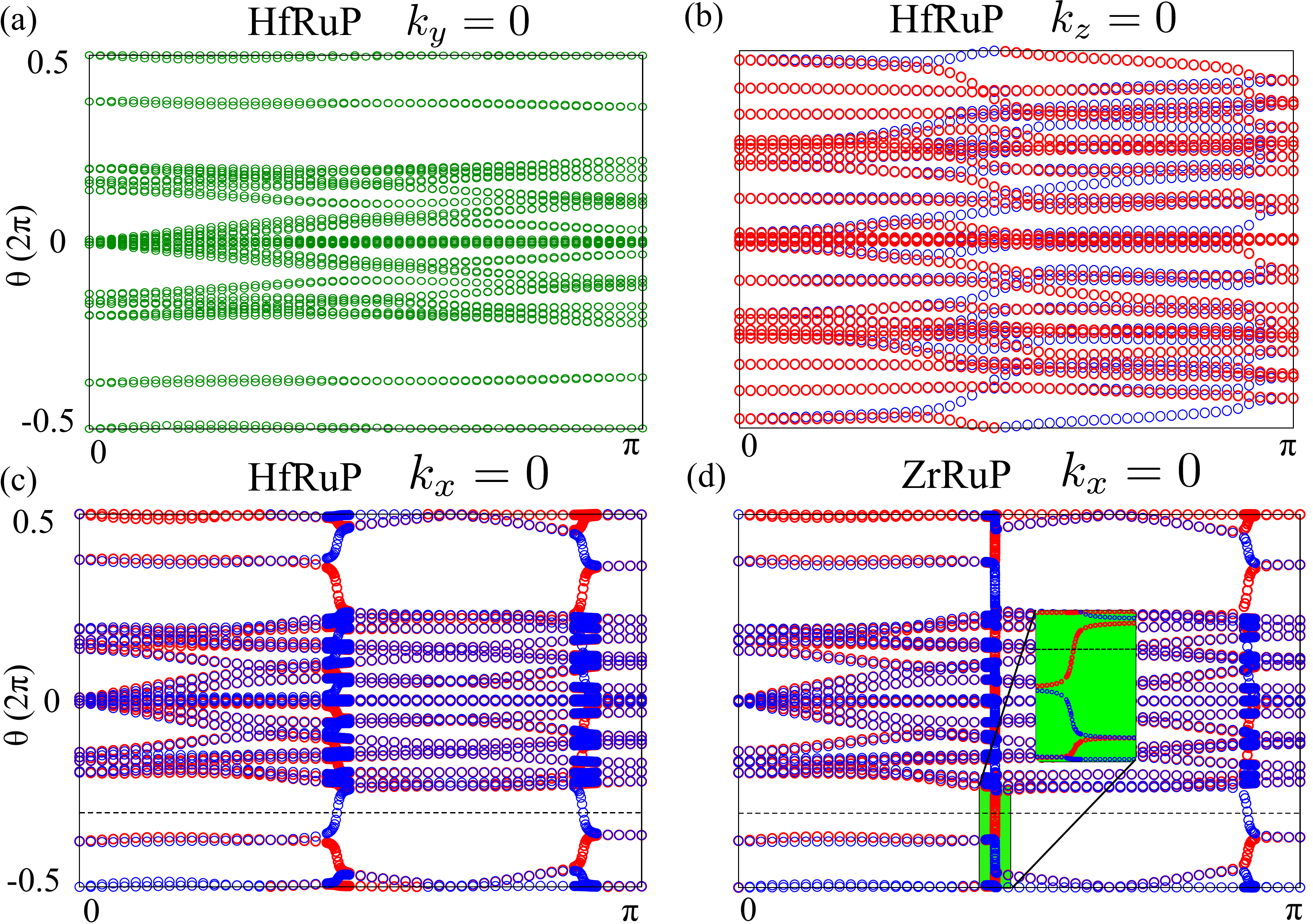}
\caption{(Color online)
The WCCs of TRI planes.
($\bold a$) and ($\bold b$) The WCCs of the $k_z$-directed and $k_y$-directed Wilson loops as a function of $k_x$ for the $k_y = 0$ and $k_z = 0$ planes, respectively.
($\bold c$) and ($\bold d$) The WCCs of the $k_z$-directed Wilson loops as a function of $k_y$ in the $k_x = 0$ plane of HfRuP and ZrRuAs, respectively.
Red and blue circles represent the flow of the WCCs for the bands with mirror $+i$ and $-i$ eigenvalues, respectively.
The horizontal dashed lines are reference lines.
}
\end{figure}

\begin{figure}[!t]
\centering
\includegraphics[angle=270,width=16 cm]{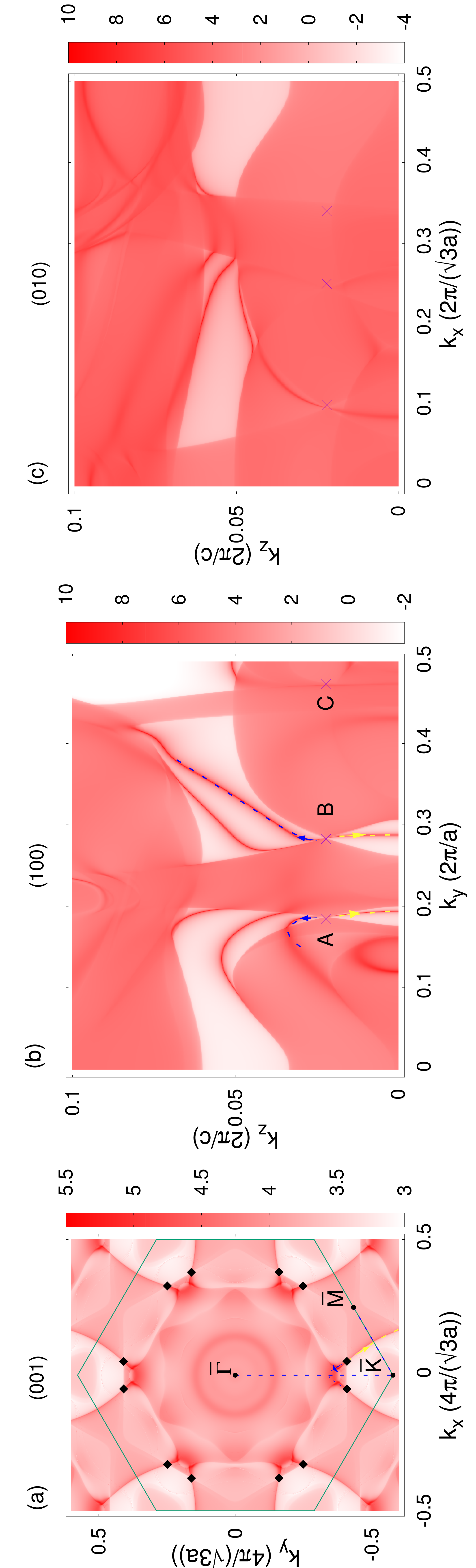}
\caption{(Color online)
The surface spectra of HfRuP. The (001)-surface ($\bold a$), (100)-surface ($\bold b$) and (010)-surface ($\bold c$) energy contours with $E-E_W=0$ meV.
The projections of the WPs are indicated by diamonds or ``x" symbols.
}
\end{figure}
\end{document}